\documentclass[a4paper]{article}
\def \be{\begin{equation}}
\def \ee{\end{equation}}
\def \bea{\begin{eqnarray}}
\def \eea{\end{eqnarray}}

\def \a{\alpha}
\def \g{\gamma}
\def \k{\kappa}

\begin{document}

\title{Colliding Plane Wave Solutions in String theory Revisited}
\author{Bin Chen\footnote{email:bchen@itp.ac.cn}\\ \\
Interdisciplinary Center of Theoretical Studies, \\
Chinese Academy of Science, P.O. Box 2735\\
Beijing 100080, P.R. China}
 \maketitle

\begin{abstract}

 We construct the colliding plane wave solutions in
the higher-dimensional gravity theory with fluxes and dilaton,
with a more general ansatz on the metric. We consider two classes
of solutions to the equations of motions and after imposing the
junction conditions we find that they are all physically
acceptable. In particular, we manage to obtain the
higher-dimensional Bell-Szekeres solutions in the Maxwell-Einstein
gravity theory, and the flux-CPW solutions in the
eleven-dimensional supergravity theory. All the solutions have
been shown to develop the late time curvature singularity.
\end{abstract}

\newpage

\section{Introduction}

The colliding plane wave solutions (CPW) have been an important
topic in the classical general relativity. They are the exact
solutions describing the collision of plane wave in a flat
background. In the four-dimensional gravity theory, the study of
CPW solutions started from the early 1970s\cite{cpw,KP}. Since
then, many exact CPW solutions have been constructed. For a
thorough review on the four dimensional CPW solutions, see
\cite{Griffiths}. One remarkable feature of CPW solutions in four
dimension is that they always develop a late time curvature
singularity\cite{Tipler,Yurtsever}. This fact could be considered
as an inevitable effect of the nonlinear gravitational focusing.
However, the initial background could play an essential role too.
In \cite{Centrella}, it has been shown that in the case that the
initial background is expanding, the focusing effect of the
nonlinear interaction between the waves could be weakened by the
expansion so that no future singularity occurs. It has been
expected that the study of the CPW solutions may tell us the
nature of the space-time singularity. Also, it has been proposed
that the gravitational plane and dilatonic waves could play an
important role in the pre-big-bang cosmology
scenarios\cite{prebigbang,Bozza}.

It is very interesting to study the colliding plane wave solutions
in the higher dimensional gravity theory. One reason is that we
may live in a higher dimensional spacetime and we have string/M
theory in 10D/11D as a candidate to describe the world. The higher
dimensional gravity theory with dilaton and various fluxes could
be taken as the low energy effective action of the string theory.
It has manifested some different features from the four
dimensional gravity theory. For example, it has been found that
the uniqueness and stability issue in higher dimensional black
holes are more subtle\cite{Kodama} and there exist the {\sl black
ring} solution with horizon topology $S^1\times S^2$ in five
dimensional gravity\cite{blackring}. Therefore, we can wish that
the study of CPW solutions reveal some new features of the
higher-dimensional gravity theory. On the other hand, due to the
existence of the dilaton and the various fluxes, the CPW solutions
in the theory have richer structure. There have been some efforts
in this direction. In \cite{Gurses1,Gurses2,Gurses3},
 the CPW solutions in the dilatonic gravity, in the
higher dimensional gravity, and in the higher dimensional
Einstein-Maxwell theory have been discussed. In \cite{Pioline},
Gutperle and Pioline tried to construct the CPW solutions in the
ten-dimensional gravity with the self-dual form flux. However,
their solutions failed to satisfy the junction condition and are
physically unacceptable.

 To
study the collinear CPW solutions in the higher dimensional
gravity, one could make a quite general ansatz
 \be
 ds^2=2e^{-M}dudv+\sum^k_{i=1}e^{A_i}dx^2_i, \hspace{5ex} (k \geq
 3)
 \ee
 where $M, A_i$ are only the function of $u,v$. Obviously, with
 such a generic ansatz, it will be very difficult to solve the equations
 of motions. In \cite{chen,chen1}, we have assumed
 $A=A_i, i=1,\cdots, n (n<k)$ and $B=A_i, i=n+1,\cdots, k$
 so that
  \be \label{metric1}
  ds^2=2e^{-M}dudv+e^A\sum_{i=1}^{n}dx_i^2+e^B\sum_{j=n+1}^{k}dy_j^2
  \ee
  to simplify the equations and
 tried to find the flux-CPW solutions with dilaton. In
 \cite{chen}, we managed to find two classes of 1-flux-CPW solutions satisfying the junction conditions:
 $(pqrw)$-type and $(f\pm g)$-type. The $(pqrw)$-type solutions
 look like the Bell-Szekeres solution, which appeared in four-dimensional Einstein-Maxwell
 gravity\cite{BS}. The $(f\pm g)$ type solutions is a new kind of
 solutions.
 Unfortunately, we also noticed
 that with metric (\ref{metric1}), it is impossible to have well-defined CPW
 solutions if we turn off the dilaton. More precisely, we found no
 higher-dimensional BS solution in Maxwell-Einstein gravity and no
 pure flux solution in eleven dimensional supergravity which has
 no dilaton. In other words, the only pure flux-CPW solution could
 only be four-dimensional BS solution. Nevertheless, with dilaton turning on,
 there always exist well-defined physical flux-CPW solution. In \cite{chen1}, we
 generalized the discussions on flux-CPW solutions to the case with
 two complementary fluxes. We found that the $(pqrw)$-type
 solution is still physically well-defined but $(f\pm g)$-type solution
 failed to satisfy the junction conditions.

In this paper, in order to investigate the flux-CPW solutions more
carefully, we go a little further and try a more complicated
ansatz:
\begin{equation}\label{metric}
ds^2=2e^{-M}dudv+e^A\sum_{i=1}^{n}dx_i^2+e^B\sum_{j=1}^{m}dy_j^2+e^C\sum_{k=1}^{l}dz_k^2.
\end{equation}
With this ansatz, we will find that the equations of motions are
still exactly solvable. And generically  we will have physically
acceptable solutions after imposing the junction conditions. We
will work on a dilatonic gravity theory with two fluxes, which
could be reduced to other cases easily. Remarkably, we notice that
the higher-dimensional pure 1-flux-CPW solutions do exist, with
the above metric ansatz.

We organize the paper as follows. In section 2, we derive the
equations of motions of the system and simplify them by changing
variables and using $(f, g)$ coordinates. In section 3, we
construct the physical CPW solutions taking into account of the
junction conditions, and discuss the future singularity issue in
these solutions. In section 4, we turn to several interesting
special cases. In particular, we consider the one-flux case, which
has been probed in \cite{chen}. We end the paper with some
discussions and conclusions.

\section{Equations of Motions and Their Reduction}

To keep the problem as generic as possible, we start from a
dilatonic gravity theory with two fluxes, whose action take the
form
\begin{equation} \label{action}
S=\int
d^Dx\sqrt{-g}\left(R-g^{\mu\nu}\partial_\mu\phi\partial_\nu\phi-\frac{1}{2(n+1)!}e^{a\phi}F^2
-\frac{1}{2(m+1)!}e^{b\phi}G^2\right).
\end{equation}
Here, $D \geq 5$ is the dimension of spacetime, $\phi$ is the
dilaton field with $a$ and $b$ being dilaton coupling constant,
and $F,G$ are two fluxes. The action could be obtained from the
low energy effective action of string/M theory with specific flux
configurations in the Einstein frame. The equations of motions are
given by

\begin{eqnarray}
R_{\mu\nu}=\partial_\mu\phi\partial_\nu\phi+\frac{1}{2n!}e^{a\phi}\left(F_{\mu\mu_1...\mu_n}F_\nu^
{\ \mu_1...\mu_n}-\frac{n}{(n+1)(m+n+l)}g_{\mu\nu}F^2\right)+\nonumber\\
+\frac{1}{2m!}e^{b\phi}\left(G_{\mu\nu_1...\nu_m}G_\nu^ {\
\nu_1...\nu_m}-\frac{m}{(m+1)(m+n+l)}g_{\mu\nu}G^2\right)
\end{eqnarray}
\begin{equation}
\partial_\mu\left(\sqrt{-g}e^{a\phi}F^{\mu\mu_1...\mu_n}\right)=0
\end{equation}
\begin{equation}
\partial_\nu\left(\sqrt{-g}e^{b\phi}G^{\nu\nu_1...\nu_m}\right)=0
\end{equation}
\begin{equation}
\frac{1}{\sqrt{-g}}\partial_\mu\left(\sqrt{-g}g^{\mu\nu}\partial_\nu\phi\right)
=\frac{a}{4(n+1)!}e^{a\phi}F^2+\frac{b}{4(m+1)!}e^{b\phi}G^2
\end{equation}

Also to keep the problem workable, we assume that the fluxes lie
along $x_i$'s and $y_j$'s respectively
 \begin{eqnarray}
   F_{ux_1...x_n}=D_u \hspace{2cm} F_{vx_1...x_n}=D_v \nonumber\\
   G_{uy_1...y_m}=E_u \hspace{2cm} G_{vy_1...y_m}=E_v.
\end{eqnarray}
In \cite{chen1}, we consider the two complimentary fluxes which
occupy the whole spacetime. Here we have additional transverse
dimensions. If we let some fluxes or dilaton field vanish, the
system can reduce to some interesting special cases, such as one
flux case, pure dilatonic case et.al, which is the subject in
section 4. One special case is the gravity coupled 1-flux without
dilaton, when we turn off the dilaton and let one flux vanish. In
eleven dimension, the study of CPW solutions in such gravity
theory may effectively describe the CPW solutions in 11D
supergravity. Recall that in the bosonic part of 11D supergravity
action, we have no dilaton but we have a $A_3\wedge F_4 \wedge
F_4$ term, where $A_3$ is the antisymmetric 3-form tensor gauge
field and $F_4$ is its field strength. However, with the above
assumption on the flux, the extra term does not contribute to the
action and the equations of motions. In this sense, the CPW
solutions in 11D supergravity could be effectively studied from
the action (\ref{action}) after reduction.

In components, the Einstein equation take the forms
\begin{equation}\label{Meq1}
nA_{uu}+mB_{uu}+lC_{uu}+M_u(nA_u+mB_u+lC_u)+\frac{1}{2}(nA_u^2+mB_u^2+lC_u^2)
=-2\phi_u^2-e^{a\phi-nA}D_u^2-e^{b\phi-mB}E_u^2
\end{equation}
\begin{equation}\label{Meq2}
nA_{vv}+mB_{vv}+lC_{vv}+M_v(nA_v+mB_v+lC_v)+\frac{1}{2}(nA_v^2+mB_v^2+lC_v^2)
=-2\phi_v^2-e^{a\phi-nA}D_v^2-e^{b\phi-mB}E_v^2
\end{equation}
\begin{eqnarray} \label{Muv}
-M_{uv}+\frac{n}{2}A_{uv}+\frac{m}{2}B_{uv}+\frac{l}{2}C_{uv}+\frac{1}{4}(nA_uA_v+mB_uB_v+lC_uC_v)
=-\phi_u\phi_v+\nonumber\\
+\frac{n-m-l}{2(n+m+l)}e^{a\phi-nA}D_uD_v
+\frac{m-n-l}{2(n+m+l)}e^{b\phi-mB}E_uE_v
\end{eqnarray}
\begin{equation}\label{Aeq}
2A_{uv}+nA_uA_v+\frac{m}{2}(A_uB_v+A_vB_u)+\frac{l}{2}(A_uC_v+A_vC_u)=-\frac{2(m+l)}{n+m+l}e^{a\phi-nA}D_uD_v
+\frac{2m}{n+m+l}e^{b\phi-mB}E_uE_v
\end{equation}
\begin{equation}\label{Beq}
2B_{uv}+mB_uB_v+\frac{n}{2}(A_uB_v+A_vB_u)+\frac{l}{2}(B_uC_v+B_vC_u)=\frac{2n}{n+m+l}e^{a\phi-nA}D_uD_v
-\frac{2(n+l)}{n+m+l}e^{b\phi-mB}E_uE_v
\end{equation}
\begin{equation}\label{Ceq}
2C_{uv}+lC_uC_v+\frac{n}{2}(A_uC_v+A_vC_u)+\frac{m}{2}(B_uC_v+B_vC_u)=\frac{2n}{n+m+l}e^{a\phi-nA}D_uD_v
+\frac{2m}{n+m+l}e^{b\phi-mB}E_uE_v.
\end{equation}
Here, as usual, the equation (\ref{Muv}) is redundant and will not
be needed in the following discussion. The equations of motions
for the dilaton and n-form, m-form potential are given by
\begin{equation}\label{Deq}
2D_{uv}+\left[a\phi-\frac{1}{2}(nA-mB-lC)\right]_uD_v+\left[a\phi-\frac{1}{2}(nA-mB-lC)\right]_vD_u=0
\end{equation}
\begin{equation}\label{Eeq}
2E_{uv}+\left[b\phi+\frac{1}{2}(nA-mB+lC)\right]_uE_v+\left[b\phi+\frac{1}{2}(nA-mB+lC)\right]_vE_u=0
\end{equation}
\begin{equation}\label{phieq}
 \phi_{uv}+\frac{1}{4}(nA+mB+lC)_u\phi_v+\frac{1}{4}(nA+mB+lC)_v\phi_u
 =\frac{a}{4}e^{a\phi-nA}D_uD_v+\frac{b}{4}e^{b\phi-mB}E_uE_v
\end{equation}

Let us introduce
\begin{equation}
U=\frac{1}{2}(nA+mB+lC) \hspace{2cm}
V=\frac{1}{2}(nA-mB-lC)\hspace{2cm}
W=\frac{1}{2}(nA-mB+lC)
\end{equation}
to rewrite the above equations. From the equations
(\ref{Aeq},\ref{Beq},\ref{Ceq}), we have
\begin{equation} \label{U}
U_{uv}+U_uU_v=0
\end{equation}
  and
\begin{eqnarray}
V_{uv}+\frac{1}{2}(U_uV_v+U_vV_u)
&=&-\frac{n(m+l)}{m+n+l}e^{a\phi-nA}D_uD_v+\frac{mn}{m+n+l}e^{b\phi-mB}E_uE_v
\label{Veq}
\\
W_{uv}+\frac{1}{2}(U_uW_v+U_vW_u)
&=&-\frac{nm}{m+n+l}e^{a\phi-nA}D_uD_v+\frac{m(n+l)}{m+n+l}e^{b\phi-mB}E_uE_v.
\label{Weq}
\end{eqnarray}
The Eq.(\ref{U}) has the well-known solution
\begin{equation}
U=\log\left[f(u)+g(v)\right],
\end{equation}
where $f,g$  are arbitrary functions, chosen usually to be
monotonic functions. One can treat $(f,g)$ as coordinates
alternative to $(u,v)$. It turns out that it is more convenient to
work in $(f,g)$ coordinates.

 Moreover, the equations (\ref{Meq1}) and (\ref{Meq2}) become
\begin{eqnarray}
U_{uu}+M_uU_u+\frac{1}{4}\left(\frac{m+n}{mn}U_u^2+\frac{l+n}{ln}V_u^2+\frac{m+l}{ml}W_u^2
+\frac{2}{n}U_uV_u-\frac{2}{m}U_uW_u-\frac{2}{l}W_uV_u\right)
=-\phi_u^2-\nonumber\\
-\frac{1}{2}e^{a\phi-nA}D_u^2-\frac{1}{2}e^{b\phi-mB}E_u^2
\end{eqnarray}
\begin{eqnarray}
U_{vv}+M_vU_v+\frac{1}{4}\left(\frac{m+n}{mn}U_v^2+\frac{l+n}{ln}V_v^2+\frac{m+l}{ml}W_v^2
+\frac{2}{n}U_vV_v-\frac{2}{m}U_vW_v-\frac{2}{l}W_vV_v\right)
=-\phi_v^2-\nonumber\\
-\frac{1}{2}e^{a\phi-nA}D_v^2-\frac{1}{2}e^{b\phi-mB}E_v^2
\end{eqnarray}
and the equations (\ref{Deq}), (\ref{Eeq}) and (\ref{phieq}) read
\bea
 2D_{uv}+(a\phi-V)_uD_v+(a\phi-V)_vD_u&=&0 \label{Deq1}\\
 2E_{uv}+(b\phi+W)_uE_v+(b\phi+W)_vE_u&=&0 \label{Eeq1}\\
 \phi_{uv}+\frac{1}{2}(U_u\phi_v+U_v\phi_u)
 &=&\frac{a}{4}e^{a\phi-U-V}D_uD_v+\frac{b}{4}e^{b\phi-U+W}E_uE_v
 \label{phieq1}
\eea

Now we can define

\begin{equation}
X=a\phi-V \hspace{1cm} Y=b\phi+W \hspace{1cm}
Z=\phi+\frac{al+(a+b)n}{4nl}V-\frac{bl+(a+b)m}{4ml}W
\end{equation}
and in terms of $(f,g)$ coordinates, the equations
(\ref{Veq},\ref{Weq},\ref{Deq1},\ref{Eeq1},\ref{phieq1}) are of
the forms \bea
(f+g)X_{fg}+\frac{1}{2}X_{f}+\frac{1}{2}X_{g}&=&\left(\frac{a^2}{4}+\frac{n(m+l)}{m+n+l}\right)
e^{X}D_{f}D_{g}+\left(\frac{ab}{4}-\frac{mn)}{m+n+l}\right)e^{Y}E_{f}E_{g}
\label{Xeq} \\
(f+g)Y_{fg}+\frac{1}{2}Y_{f}+\frac{1}{2}Y_{g}&=&\left(\frac{ab}{4}-\frac{mn}{m+n+l}\right)
e^{X}D_{f}D_{g}+\left(\frac{b^2}{4}+\frac{m(n+l))}{m+n+l}\right)e^{Y}E_{f}E_{g}
\label{Yeq}\\
 2D_{fg}+X_{f}D_{g}+X_{g}D_{f}&=&0 \label{Deq2}\\
2E_{fg}+Y_{f}E_{g}+Y_{g}E_{f}&=&0 \label{Eeq2}\\
(f+g)Z_{fg}+\frac{1}{2}(Z_f+Z_g)&=&0 \label{Zeq}
 \eea
 There exist a large class of solutions of the equation (\ref{Zeq})\cite{Griffiths,BGM}, which is of
 the form of the Euler-Darboux equation. We are not
 going to discuss all these solutions here. Instead, we just focus on the well-known
 Khan-Penrose-Szekeres
 solution:
\begin{equation} \label{Z}
Z=\kappa_1\log\frac{w-p}{w+p}+\kappa_2\log\frac{r-q}{r+q}.
\end{equation}
The equations (\ref{Xeq},\ref{Yeq},\ref{Deq2},\ref{Eeq2}) are
coupled differential equations, which may be taken as a
generalized Ernst equations. We will make ansatz and solve these
equations in the next section.

The equations (\ref{Meq1},\ref{Meq2}) can also be simplified to
\bea
 S_f+(f+g)\left[\frac{V_f^2}{4n}+\frac{W_f^2}{4m}+\frac{(V_f-W_f)^2}{4l}+\phi^2_f\right]
 +\frac{1}{2}e^XD_f^2+\frac{1}{2}e^YE_f^2&=&0 \label{Sf} \\
 S_g+(f+g)\left[\frac{V_g^2}{4n}+\frac{W_g^2}{4m}+\frac{(V_g-W_g)^2}{4l}+\phi^2_g\right]
 +\frac{1}{2}e^XD_g^2+\frac{1}{2}e^YE_g^2&=&0 \label{Sg}
 \eea
 by introducing
\begin{equation}
S=M-(1-\delta)\log(f+g)+\log(f_{u}g_{v})+\frac{1}{2n}V-\frac{1}{2m}W
\end{equation}
with
 \be
 \delta=\frac{m+n}{4mn}.
 \ee

 Once we solve the $X,Y,Z,D,E$, we can use the relation
 \bea
 \phi&=&\frac{4mnlZ+m(al+(a+b)n)X+n(bl+(a+b)m)Y}{\a}\\
 V&=&\frac{4mnl(aZ-X)+n(bl+(a+b)m)(aY-bX)}{\a}\\
 W&=&\frac{4mnl(Y-bZ)+m(al+(a+b)n)(aY-bX)}{\a}
 \eea
 to get $\phi,V,W$, where
 \be
 \a=mn(a+b)^2+l(ma^2+nb^2+4mn).
 \ee
 With $\phi,V,W$, one could integrate (\ref{Sf},\ref{Sg})
 to get $S$ and then $M$ so as to obtain all the metric
 components. At the first looking, the  three relations on $\phi, V,W$ are
 quite involved and the integration on $S$ seems to be a forbidden
 task. Nevertheless, with the solutions we will discuss in this paper, we
 succeed in getting the exact metric components.

\section{Physical CPW Solutions}

In the study of the collision of the gravitational plane waves,
one usually divides the spacetime into four regions: past
P-region($u<0,v<0$), right R-region($u>0, v<0$), left
L-region($u<0, v>0$) and future F-region($u>0,v>0$), which
describes the flat Minkowski spacetime, the incoming waves from
right and left, and the colliding interaction region respectively.
The general recipe to construct the CPW solutions is to solve the
equations of motions in the forward region and then reduce the
solutions to other regions, requiring the metric to be continuous
and invertible in order to paste the solutions in different
regions. More importantly, one need to impose the junction
conditions to get an  acceptable physical solution. In this
section, we try to find the solutions to the equations of motions
in the forward region and then reduce them to other regions and
impose the junction conditions to get the physical CPW solutions.

\subsection{Solutions to the Equations of Motions}

As we mentioned, there exist a large class of solutions to the
Euler-Darboux equation (\ref{Zeq}). And for the coupled
differential equations (\ref{Xeq}-\ref{Eeq2}), if we assume that
$X_f=Y_f, X_g=Y_g$ and $D \propto E$, then the equations could
reduce to the ones in 1-flux case\cite{chen}, which are related to
the Ernst equation. We will focus on two kinds of solutions of
(\ref{Xeq}-\ref{Eeq2}): one is $(pqrw)$-type and the other is $(f
\pm g)$-type.

\begin{itemize}
\item   (pqrw)-type (BS type) solution:\\

Let us make the following ansatz:
\begin{eqnarray}
X=-\log c_{1}\frac{rw+pq}{rw-pq} \hspace{2cm}
Y=-\log c_{2}\frac{rw+pq}{rw-pq}\label{X1}\\
D=\gamma_{1}\cdot(pw-rq)\hspace{2.5cm}
E=\gamma_{2}\cdot(pw-rq)\label{D1}
\end{eqnarray}
 where
\begin{equation}
 p:=\sqrt{\frac 12-f} \hspace{1cm}
 q:=\sqrt{\frac 12-g} \hspace{1cm}
 r:=\sqrt{\frac 12+f} \hspace{1cm}
 w:=\sqrt{\frac 12+g}.
\end{equation}

Note that there seem to be two free parameters $c_1, c_2$ in $X,
Y$, however, the continuity condition on the metric in four
patches restrict $c_1=c_2=1$. And the equations
(\ref{Xeq},\ref{Yeq}) require
 \bea
 \g_1^2&=&2\cdot\frac{b(b-a)(m+n+l)+4m(2n+l)}{\a}
 \\
 \g_2^2&=&2\cdot\frac{a(a-b)(m+n+l)+4n(2m+l)}{\a}
 \eea

After a straightforward but tedious calculation, we obtain
 \be
 S_f+(f+g)\left[\frac{4mnl}{\a}Z_f^2+\frac{\g^2_1+\g^2_2}{8}X_f^2\right]+\frac{1}{2}e^XD^2_f+\frac{1}{2}e^YE^2_f=0
 \ee
 and the similar relation to $S_g$. After integration over $f$ and
 $g$, one has
 \bea
 S&=& b_1\log(1-2f)(1+2g)+b_2\log(1+2f)(1-2g) \nonumber \\
 & & +b_3\log(f+g)+b_4\log(1+4fg+\sqrt{(1-4f^2)(1-4g^2)})
 \label{S1}
 \eea
 where
 \bea
 b_1&=&\frac{\g^2_1+\g^2_2}{8}+\frac{4mnl\k_1^2}{\a} \label{b1a}\\
 b_2&=&\frac{\g^2_1+\g^2_2}{8}+\frac{4mnl\k_2^2}{\a} \label{b2a}\\
 b_3&=&-\frac{\g^2_1+\g^2_2}{8}-\frac{4mnl}{\a}(\k_1+\k_2)^2 \label{b3a}\\
 b_4&=&\frac{8mnl\k_1\k_2}{\a}
 \eea
and then the metric and the dilaton is given by
 \begin{eqnarray}
e^{-M}&=&f_{u}g_{v}
\left[(1-2f)(1+2g)\right]^{-b_1}\left[(1+2f)(1-2g)\right]^{-b_2}(f+g)^{-b_3-1+\delta}\nonumber\\
& &\times\left[1+4fg+4pqrw\right]^{-b_4}
\left(\frac{rw+pq}{rw-pq}\right)^{s_1}
\left[\left(\frac{w-p}{w+p}\right)^{\kappa_1}\left(\frac{r-q}{r+q}\right)^{\kappa_2}\right]^{\frac{2l(ma+nb)}{\alpha}} \label{M1}\\
e^{nA}&=&(f+g)\left[\left(\frac{w-p}{w+p}\right)^{\kappa_1}
\left(\frac{r-q}{r+q}\right)^{\kappa_2}\right]^\frac{4mnla}{\alpha}
\left(\frac{rw+pq}{rw-pq}\right)^{s_2} \label{A1}\\
e^{mB}&=&(f+g)\left[\left(\frac{w-p}{w+p}\right)^{\kappa_1}
 \left(\frac{r-q}{r+q}\right)^{\kappa_2}\right]^\frac{4mnlb}{\alpha}
 \left(\frac{rw+pq}{rw-pq}\right)^{s_3} \label{B1}\\
 e^{lC}&=&\left[\left(\frac{w-p}{w+p}\right)^{\kappa_1}
 \left(\frac{r-q}{r+q}\right)^{\kappa_2}\right]^\frac{-4mnl(a+b)}{\alpha}
 \left(\frac{rw+pq}{rw-pq}\right)^{s_4} \label{C1}\\
 e^\phi &=&\left[\left(\frac{w-p}{w+p}\right)^{\kappa_1}
 \left(\frac{r-q}{r+q}\right)^{\kappa_2}\right]^\frac{4mnl}{\alpha}
 \left(\frac{rw+pq}{rw-pq}\right)^{s_5}
\end{eqnarray}
where
 \bea
 s_1&=&\frac{l(a-b)^2+(b^2-a^2)(m-n)+4l(m+n)}{2\alpha} \\
 s_2&=&{\frac{n}{\alpha}[(b-a)(bl+(b+a)m)+4ml]} \\
 s_3&=&{\frac{m}{\alpha}[(a-b)(al+(b+a)n)+4nl]} \\
 s_4&=&{-\frac{l}{\alpha}[(a-b)(ma-nb)+8mn]} \\
 s_5&=&-\frac{1}{\alpha}[l(ma+nb)+2mn(a+b)]
 \eea

Therefore we have a two-parameter family of solution labelled by
$\k_1,\k_2$.

\item $(f\pm g)$-type solution

As the flux-CPW solution with  metric ansatz (\ref{metric1})
discussed in \cite{chen,chen1}, we may assume that $X,Y$ is the
function of $(f+g)$ and $D,E$ is the function of $(f-g)$. The
equations (\ref{Deq2},\ref{Eeq2}) tell us
\begin{equation}\label{D2}
D=\gamma_1\cdot(f-g) \hspace{2cm}  E=\gamma_2\cdot(f-g)
\end{equation}
for some constants $\gamma_1,\gamma_2$.  And (25), (26) can be
solved by
 \bea
X&=&-\log\left[\frac{(f+g)}{\a_1}\cosh^2\left(c_1\log{\frac{c_2}{f+g}}\right)\right] \label{X2}\\
Y&=&-\log\left[\frac{(f+g)}{\a_2}\cosh^2\left(c_1\log{\frac{c_2}{f+g}}\right)\right]
 \eea
 where
 \bea
 \a_1&=&\frac{2c_1^2[b(b-a)(m+n+l)+4m(2n+l)]}{\a \g_1^2} \\
 \a_2&=&\frac{2c_1^2[a(a-b)(m+n+l)+4n(2m+l)]}{\a \g_2^2}
 \eea

After integrating $f$ and $g$, one can read \bea
 S&=& b_1\log(1-2f)(1+2g)+b_2\log(1+2f)(1-2g)+b_3\log(f+g) \nonumber \\
 & & +b_4\log(1+4fg+\sqrt{(1-4f^2)(1-4g^2)})+b_5\log
 \cosh\left(c_1\log{\frac{c_2}{f+g}}\right) \label{S2}
 \eea
 where
 \bea
 b_1&=&\frac{4mnl\k_1^2}{\a} \label{b1b}\\
 b_2&=&\frac{4mnl\k_2^2}{\a} \label{b2b}\\
 b_3&=&-\frac{(\a_1\g^2_1+\a_2\g^2_2)(4c_1^2+1)}{8c_1^2}-\frac{4mnl}{\a}(\k_1+\k_2)^2 \label{b3b}\\
 b_4&=&\frac{8mnl\k_1\k_2}{\a}\\
 b_5&=&-\frac{\a_1\g^2_1+\a_2\g^2_2}{2c_1^2}.
 \eea
And the metric components and the dilaton follow
 \bea
e^{-M}&=&t_0f_ug_v\left[(1-2f)(1+2g)\right]^{-b_1}\left[(1+2f)(1-2g)\right]^{-b_2}(f+g)^{-b_3-1+\delta}
 \left[\cosh\left(c_1\log\frac{c_2}{f+g}\right)\right]^{-b_5}\nonumber\\
 & &\times
 \left[(f+g)\cosh^2\left(c_1\log\frac{c_2}{f+g}\right)\right]^{s_1}
 \left[1+4fg+4pqrw\right]^{-b_4}
\left[\left(\frac{w-p}{w+p}\right)^{\kappa_1}\left(\frac{r-q}{r+q}\right)^{\kappa_2}\right]^
{-\frac{2l(ma+nb)}{\alpha}} \label{M2} \\
 e^{nA}&=&t_1(f+g)
 \left[(f+g)\cosh^2\left(c_1\log\frac{c_2}{f+g}\right)\right]^{s_2}
\left[\left(\frac{w-p}{w+p}\right)^{\kappa_1}\left(\frac{r-q}{r+q}\right)^{\kappa_2}\right]^{\frac{4mnla}{\alpha}} \label{A2}\\
 e^{mB}&=&t_2(f+g)
\left[(f+g)\cosh^2\left(c_1\log\frac{c_2}{f+g}\right)\right]^{s_3}
\left[\left(\frac{w-p}{w+p}\right)^{\kappa_1}\left(\frac{r-q}{r+q}\right)^{\kappa_2}\right]^\frac{4mnlb}{\alpha} \label{B2}\\
e^{lC}&=&t_1^{-1}t_2^{-1}\left[(f+g)\cosh^2\left(c_1\log\frac{c_2}{f+g}\right)\right]^{s_4}
\left[\left(\frac{w-p}{w+p}\right)^{\kappa_1}\left(\frac{r-q}{r+q}\right)^{\kappa_2}\right]^\frac{-4mnl(a+b)}{\alpha} \label{C2}\\
 e^\phi &=& t_3\left[(f+g)\cosh^2\left(c_1\log\frac{c_2}{f+g}\right)\right]^{s_5}
\left[\left(\frac{w-p}{w+p}\right)^{\kappa_1}\left(\frac{r-q}{r+q}\right)^{\kappa_2}\right]^{\frac{4mnl}{\alpha}}
 \eea
 where $s_i$'s are the same parameters above
and
\bea
 t_0&=&\a_1^{-\frac{4ml+b((b-a)l+(a+b)(m-n))}{2\a}}\a_2^{-\frac{4nl-a((b-a)l+(a+b)(m-n))}{2\a}} \nonumber\\
 t_1&=&\a_1^{-n\frac{4ml+b(bl+(a+b)m)}{\a}}\a_2^{na\frac{bl+(a+b)m}{\a}} \nonumber\\
 t_2&=&\a_1^{mb\frac{al+(a+b)n}{\a}}\a_2^{-m\frac{4nl+a(al+(a+b)n)}{\a}} \nonumber\\
 t_3&=&\a_1^{-m\frac{al+(a+b)n}{\a}}\a_2^{-n\frac{bl+(a+b)m}{\a}}
 \nonumber
\eea After imposing the continuity condition on the metric, one
 get quite involved constraints on $c_1, c_2,\a_1,\a_2$. One
can choose $c_2=0$ and simplify the constraints to be
$\a_1=\a_2=1$, which lead to $t_i=1$ for $i=0,\cdots,3$ and the
relations among $c_1^2$ and $\g_1, \g_2$. In short, we have a
three-parameter family of solutions, labelled by $\k_1,\k_2$ and
$c_1$.

\end{itemize}

Up to now, we  have solved the equations of motions in the
F-region. Actually one can reduce the above solutions to the ones
for the L-region, the R-region, and the P-region if one do the
following replacements:
\begin{equation} \label{f}
f(u)=f_0\hspace{1cm}f_u(1-2f)^{-b_1}\mid_{f=f_0}=-1\hspace{2cm}
\mbox{for}\
 \ \ u<0
\end{equation}
\begin{equation} \label{g}
g(v)=g_0\hspace{1cm}g_v(1-2g)^{-b_2}\mid_{g=g_0}=-1\hspace{2cm}
\mbox{for}\
 \ \ v<0
\end{equation}
      where
$b_1, b_2$ have been given in (\ref{b1a},\ref{b1b}),
(\ref{b2a},\ref{b2b}) and $f_0$, $g_0$ are constants. Taking into
account of the continuous and invertible conditions on the metric,
we are able to fix the values of
\begin{equation}
f_0=g_0=1/2.
\end{equation}
Actually, in order to simplify the expressions, we have use the
continuous and invertible condition to fix the value of $c_i$'s
and $\a_i$'s.

The metric (\ref{metric}) is in the Rosen coordinates. In order to
show that the incoming waves looks like the plane-waves, it is
more convenient to write the metric in Brinkmann coordinates. In
the incoming region, e.g. R-region ($v<0$), the metric in the
Brinkmann coordintes takes the form:
 \bea
 ds^2&=&2dx^+dx^-+\left(H_x(x^+)\sum_{i=1}^nX^2_i+H_y(x^+)\sum_{j=1}^mY^2_j+H_z(x^+)\sum_{k=1}^lZ^2_k\right)(dx^+)^2
 \nonumber \\
 & &+\sum_{i=1}^ndX^2_i+\sum_{j=1}^mdY^2_j+\sum_{k=1}^ldZ^2_k,
 \eea
where $x^+$ is related to $u$ through
 \be
 e^{-M}du = dx^+
 \ee
and
 \bea
 H_x&=&e^{-A}\frac{d^2e^A}{dx^{+2}}=e^{2M}(A_{uu}+M_uA_u+A^2_u)
 \nonumber \\
H_y&=&e^{-B}\frac{d^2e^B}{dx^{+2}}=e^{2M}(B_{uu}+M_uB_u+B^2_u)
 \nonumber \\
H_z&=&e^{-C}\frac{d^2e^C}{dx^{+2}}=e^{2M}(C_{uu}+M_uC_u+C^2_u)
 \nonumber
 \eea
It is straightforward to write down the explicit Brinkmann form of
the metric corresponding to different CPW solutions.

\subsection{Imposing the Junction Conditions}

The junction conditions are essential to make the solutions
obtained be physically acceptable. The detailed discussions on the
junction conditions can be found in \cite{chen}. The key points
are
\begin{itemize}
\item[{\it (1)}] The metric must be continuous and invertible;
\item[{\it (2)}] If the metric is $C^1$, then impose the Lichnerowicz condition: the metric has to be at least $C^2$.
Otherwise, if the metric is piecewise $C^1$, then impose the OS
junction conditions\cite{OS} which require
    \begin{equation}
    g_{\mu\nu},\hspace{1cm}
    \sum_{ij}g^{ij}g_{ij,0},\hspace{1cm}
    \sum_{ij}g^{i0}g_{ij,0},\hspace{2cm}(i,j\neq0).
    \end{equation}
to be continuous across the null surface (note that ``$0$" in the
above formulae stands for $u=0$ or $v=0$). From our ansatz on the
metric, the OS condition means that $U,V,M$ need to be continuous
and $U_u =0$ across the junction at $u=0$. The same happens at the
junction $v=0$.
\item[{\it (3)}]The curvature invariants $R$, $R_2$ do not blow up at the junction.
\end{itemize}
The first condition is natural to paste the solutions in different
regions together. The second condition comes from the requirement
that the stress tensor could be piecewise continuous instead of
being continuous, namely the stress tensor may have finite jump
but not $\delta$-function jump across the junction, such that the
Ricci tensor is allowed to be piecewise continuous. The third
condition is from the requirement that the curvature invariants
$R$ and $R_2$ should not have poles at the junction.

In order to study the behavior of the solution near the junction,
we assume that near junction,
\begin{equation} \label{f}
f(u)=f_0(1-d_1u^{n_1})\hspace{2cm}u\sim0^+
\end{equation}
\begin{equation} \label{g}
g(v)=g_0(1-d_2v^{n_2})\hspace{2cm}v\sim0^+
\end{equation}
where $n_i, i=1,2$ are the boundary exponents.

From the continuous condition on the metric and conditions
(\ref{f},\ref{g}), one has \be f_0=g_0 = 1/2 \ee and \be
b_i=1-\frac{1}{n_i},\hspace{2cm}d_i=\left(\frac{2}{n_i}\right)^{n_i},
\hspace{5ex} i=1,2. \ee

With the expansion (\ref{f},\ref{g}), it is not hard to figure out
the near junction behavior of the metric components. Near $u\sim
0$ (same for $v\sim 0$), the most singular terms in the metric
are:
 \bea
 M_u &\sim & \left(
 u^{\frac{n_1}{2}-1}e_0(\nu)+l.s.t.\right)\Theta(u)\\
 A_u &\sim & \left(
 u^{\frac{n_1}{2}-1}e_1(\nu)+l.s.t.\right)\Theta(u)\\
 B_u &\sim & \left(
 u^{\frac{n_1}{2}-1}e_2(\nu)+l.s.t.\right)\Theta(u)\\
 C_u &\sim & \left(
 u^{\frac{n_1}{2}-1}e_3(\nu)+l.s.t.\right)\Theta(u)
 \eea
 where $e_i(v), i=0,\cdots, 3$ are some nonzero functions of $v$ and $\Theta(u)$ is the Heaviside step function.
 \footnote{Actually, the detailed discussions on near junction
 behavior are quite similar to the ones in \cite{chen, chen1} so
 we just give out the final answer to save the space.} After
 imposing the condition (2) above, we have
 \begin{equation}
\left\{\begin{array}{l}(1)\ \ 1<n_i\leq 2\hspace{3cm}\mbox{metric
is piecewise $C^1$ }\\(2)\ \ n_i>2\hspace{3.6cm}\mbox{metric is at
least piecewise $C^2$ }\end{array}\right.
\end{equation}

As the last step, we have to impose the condition (3) on $R,R_2$.
Now, one has
 \be
 R=2e^M\phi_u\phi_v+\frac{m-n+l}{m+n+l}D_uD_v\frac{e^{M+X}}{f+g}
 +\frac{n-m+l}{m+n+l}E_uE_v\frac{e^{M+Y}}{f+g}
 \ee
 from the Einstein equations and also
 \be
 R_2 = 2e^{2M}R_{uv}^2 + 2e^{2M}R_{uu} R_{vv} + n e^{-2A}
R_{xx}^2 + me^{-2B} R_{yy}^2+le^{-2C}R_{zz}^2.
 \ee
 From the near
junction behavior of the metric and fields, one finds
 that near $u\sim 0$ (same for $v\sim 0$),
 \be
 R\sim u^{\frac{n_1}{2}-1} \sim R_2
 \ee
In order to keep $R, R_2$ from blowing up, one needs to ask $n_i
\geq 2$.

Taking into account  the constraints from all the junction
conditions, we have the following physical possibilities :
 \be
b_i=1-\frac{1}{n_i}
 \ee and
\begin{equation}
\left\{\begin{array}{l}(1)\ n_i=2\hspace{3cm}\mbox{metric is
piecewise $C^1$}\\(2)\ n_i>2\hspace{3cm}\mbox{metric is at least
piecewise $C^2$}\end{array}\right.
\end{equation}

This leads to the following relations
 \be
 \frac{1}{2} \leq b_i \leq 1, \hspace{5ex}i=1,2
 \ee

Let us turn to our solutions in the above section. From
(\ref{b1b},\ref{b2b}), it is obvious that we have physical
acceptable solutions since there are two free parameters
$\k_1,\k_2$. Therefore, physical $(f\pm g)$ solutions always
exist. As for the $(pqrw)$ type solution, from
(\ref{b1a},\ref{b2a}), the only trouble may comes from $b_i >1$,
which could happen when
 \be
 \frac{\g_1^2+\g_2^2}{8} > 1.
 \ee
This is possible. One example is when $a=-b$ and $l=1$, which lead
to
 \be
 \frac{\g_1^2+\g_2^2}{8}=\frac{a^2(m+n+1)+4mn+m+n}{a^2(m+n)+4mn}.
 \ee
Since $m,n$ are integers, the above relation must be bigger than
$1$. It seems that there may not exist physically acceptable
$(pqrw)$-type solution. Certainly, a case-by-case study is needed.
One can hope that in general, there could exist $(pqrw)$ type
solutions.

Next, we would like to consider the future singularity issue. In
the various four-dimensional CPW solutions and higher-dimensional
CPW solutions, it has been found that the future singularity is
always developed\cite{Tipler,Yurtsever,chen,chen1}. We now show
that it is the case here. Define a hyper-surface $S_0$:
\begin{equation}
f(u)+g(v)=0
\end{equation}
near which the metric may blow up or vanish. The singular behavior
of the metric near $S_0$ is easy to read out. From the fact that
 \be
 e^{lC}=e^{2U-nA-mB}=(f+g)^2e^{-nA}e^{-mB},
 \ee
 we know that it is impossible to require the metric to be regular and
 invertible near $S_0$. In other words, the metric must be
 singular at $S_0$. To see that this singularity
 is not a coordinate singularity, let us check the curvature
 invariants $R, R_2, R_4$, whose most singular terms near $S_0$ all take
 the form:
 \be
 e^{2M}(f+g)^{-4}\sim (f+g)^{2\beta}
 \ee
 where
 \be
 \beta=b_3-1-\delta-s_1(1-2(1-\epsilon)c_1)-b_5c_1-(\k_1+\k_2)\frac{2l(ma+nb)}{\a}.
 \ee
For the $(pqrw)$-type solution, with some efforts, one can get
 \be
 \beta\leq -1-\frac{1}{\a}(na^2+mb^2+l(a-b)^2+(4mn+3ml+3nl)) <0
 \ee
 so the future curvature singularity will be developed. As for
 the $(f\pm g)$-type solution, the analysis on $\beta$ is quite
 involved but we believe that the same conclusion could be
 reached.

\section{Some Special cases}

In the above sections, we have worked on a quite general higher
dimensional gravity theory with two fluxes and dilaton. The key
point in this paper is the metric ansatz (\ref{metric}). One
interesting issue is to revisit the flux-CPW solutions in various
special cases, with the metric (\ref{metric}). We will find that
the CPW solutions in some cases could be reduced directly from the
solutions we have obtained in the above section, while in other
cases, we have to solve the reduced equations of motions
independently.

\subsection{Case I: One-flux CPW solutions:}

The first case is to set $b=0, E=0$ such that our theory reduce to
the one-flux dilatonic gravity, which has been discussed
thoroughly in \cite{chen} under a different metric ansatz
(\ref{metric1}). In \cite{Gurses1}, the CPW solutions of the
four-dimensional Einstein-Maxwell-dilaton theory has been
constructed. In our higher dimensional case, we have
 \bea
 X&=&a\phi-V \nonumber \\
 Y&=& W \nonumber \\
 Z&=& \phi+\frac{a}{4nl}((l+n)V-nW). \nonumber
 \eea
The equations on $X,Y,Z,D$ are easily reduced from
(\ref{Xeq},\ref{Yeq},\ref{Zeq},\ref{Deq2}). It is easy to find out
the solutions to $X,D,Z$. However, in order to solve $Y$, one
needs to introduce
 \be
 K=\frac{mn}{m+n+l}X +
 \left(\frac{a^2}{4}+\frac{n(m+l)}{m+n+l}\right)Y
 \ee
 which satisfy
 \be
 (f+g)K_{fg}+\frac{1}{2}(K_f+K_g)=0
 \ee
 with the solution chosen to be
 \be \label{K}
 K=\kappa_3\log\frac{w-p}{w+p}+\kappa_4\log\frac{r-q}{r+q}.
 \ee
Therefore, from the solutions $K,X$, one can get the solutions of
$Y$.

In terms of $Z,K,X$, we get
 the equation on $S_f$(same to $S_g$)
 \bea
 0&=&S_f+(f+g)\left[\frac{4nlZ_f^2}{4ln+a^2(l+n)}+\frac{(m+n+l)X^2_f}{a^2(m+n+l)+4n(m+l)}\right.\nonumber\\
 & &\left.  +\frac{4(n+m+l)^2K^2_f}{[4ln+a^2(l+n)][a^2(m+n+l)+4n(m+l)]m}\right]+\frac{1}{2}e^X D_f^2
 \eea

In this case, the solutions to $X,D$ could be $(pqrw)$-type as in
(\ref{X1},\ref{D1}) with
 \be
 c_1=1, \hspace{5ex}\g_1^2=\frac{8(m+n+l)}{a^2(m+n+l)+4n(m+l)}
 \ee
 or $(f\pm g)$-type as in (\ref{X2},\ref{D2}) with
 \be
 \frac{\g_1^2}{c_1^2}=\frac{8(m+n+l)}{a^2(m+n+l)+4n(m+l)}.
 \ee
 Then in a straightforward way we get the same form of $S$ as
(\ref{S1},\ref{S2}) but with
 \bea
 b_1&=&\frac{4nlm[a^2(m+n+l)+4n(m+l)]\k_1^2+4(n+m+l)^2\k^2_3+ \epsilon m(m+n+l)[4ln+a^2(l+n)]}
 {[4ln+a^2(l+n)][a^2(m+n+l)+4n(m+l)]m}\\
 b_2&=&\frac{4nlm[a^2(m+n+l)+4n(m+l)]\k_2^2+4(n+m+l)^2\k^2_4+ \epsilon m(m+n+l)[4ln+a^2(l+n)]}
 {[4ln+a^2(l+n)][a^2(m+n+l)+4n(m+l)]m}  \\
 b_3&=&-\frac{4nl(\k_1+\k_2)^2}{4ln+a^2(l+n)}-\frac{4(n+m+l)^2(\k_3+\k_4)^2}
 {[4ln+a^2(l+n)][a^2(m+n+l)+4n(m+l)]m}-\frac{(m+n+l)(4(1-\epsilon) c_1^2+1)}{a^2(m+n+l)+4n(m+l)} \nonumber \\
 b_4&=&\frac{8nlm[a^2(m+n+l)+4n(m+l)]\k_1\k_2+8(n+m+l)^2\k_3\k_4}{[4ln+a^2(l+n)][a^2(m+n+l)+4n(m+l)]m} \nonumber \\
 b_5&=&(\epsilon-1)\frac{4(m+n+l)}{a^2(m+n+l)+4n(m+l)} \nonumber .
 \eea
Here
 \be
 \epsilon=\left\{\begin{array}{l}
 1, \hspace{5ex} \mbox{for $(pqrw)$-type solution} \\
 0, \hspace{5ex} \mbox{for $(f\pm g)$-type solution}
 \end{array}\right.
 \ee

Similarly, we have metric components and dilaton field as
 \bea
 e^{-M}&=&f_{u}g_{v}
\left[(1-2f)(1+2g)\right]^{-b_1}\left[(1+2f)(1-2g)\right]^{-b_2}(f+g)^{-b_3-1+\delta}\nonumber\\
&
&\times\left[1+4fg+4pqrw\right]^{-b_4}\left[\cosh\left(c_1\log\frac{c_2}{f+g}\right)\right]^{-b_5}
e^{-s_1X+a_1K}\exp\left\{\frac{2laZ}{4ln+a^2(l+n)}\right\})
 \label{M3}\\
e^{nA}&=&(f+g) e^{-s_2X+a_2K}\exp\left\{\frac{4nlaZ}{4ln+a^2(l+n)}\right\}\label{A3}\\
e^{mB}&=&(f+g)e^{-s_3X+a_3K} \label{B3}\\
 e^{lC}&=&e^{-s_4X+a_4K}\exp\left\{\frac{-4nlaZ}{4ln+a^2(l+n)}\right\}\label{C3}\\
 e^\phi &=&e^{-s_5X+a_5K}\exp\left\{\frac{4nlZ}{4ln+a^2(l+n)}\right\}
 \eea
where

 \bea
 s_1&=\frac{2(m-n+l)}{a^2(m+n+l)+4n(m+l)},
 \hspace{5ex}a_1&=-\frac{2(m+n+l)[(m-l-n)a^2-4ln]}{[4ln+a^2(l+n)][a^2(m+n+l)+4n(m+l)]m}
 \\
 s_2&=\frac{4n(m+l)}{a^2(m+n+l)+4n(m+l)},
 \hspace{5ex}a_2&=\frac{4(m+n+l)a^2n}{[4ln+a^2(l+n)][a^2(m+n+l)+4n(m+l)]}\\
 s_3&=\frac{-4mn}{a^2(m+n+l)+4n(m+l)},
 \hspace{5ex}a_3&=-\frac{4(m+n+l)}{a^2(m+n+l)+4n(m+l)} \\
 s_4&=\frac{-4nl}{a^2(m+n+l)+4n(m+l)},
 \hspace{5ex}a_4&=\frac{4(m+n+l)l(a^2+4n)}{[4ln+a^2(l+n)][a^2(m+n+l)+4n(m+l)]} \\
 s_5&=-\frac{a(m+n+l)}{a^2(m+n+l)+4n(m+l)},
 \hspace{5ex}a_5&=\frac{4an(m+n+l)}{[4ln+a^2(l+n)][a^2(m+n+l)+4n(m+l)]}
 \eea
 and
 \be
 e^{-X}=\left\{\begin{array}{ll}
 \frac{rw+pq}{rw-pq}, \hspace{5ex}&\mbox{for $(pqrw)$-type
 solution}\\
 (f+g)\cosh^2\left(c_1\log\frac{c_2}{f+g}\right) \hspace{5ex}&\mbox{for $(f\pm g)$-type
 solution}
 \end{array} \right.
 \ee
and $K$ is of the form (\ref{K}), $Z$ is of the form (\ref{Z}).

In short, we have a four-parameter family of the $(pqrw)$-type
solutions labelled by $\k_1,\k_2,\k_3,\k_4$ and a five-parameter
family of the $(f\pm g)$-type solutions labelled by
$\k_1,\k_2,\k_3,\k_4$ and $c_1$.

In order to get the physical solutions, we impose the junction
conditions which similarly leads to $n_i \geq 2$, or equivalently,
$1 \geq b_i \geq 1/2$, for $i=1,2$. As before, the $(f\pm g)$-type
solution exist anyway. The only trouble in $(pqrw)$-type solutions
may come from
 \be
 \frac{m+n+l}{a^2(m+n+l)+4n(m+l)} >1.
 \ee
It is not difficult to see that this is impossible since the
L.H.S. of the above relation is always less than 1. Therefore, in
the one-flux case, we can always have two classes of dilatonic
flux-CPW solutions.

In the same way, one can study the future singularity of these
solutions. Near $S_0: f+g=0$, the metric is always singular. And
the most singular terms in the curvature invariants are all  of
the form $(f+g)^{2\beta}$ with
 \be
 \beta=b_3-1-\delta-s_1(1-2(1-\epsilon)c_1)-b_5c_1-(\k_1+\k_2)\frac{2la}{4ln+a^2(l+n)}.
 \ee
We have proved that $\beta$ is negative for both types of
solutions. This indicates that the future curvature singularity is
unavoidable.

Furthermore, if one turn off the dilaton, one gets a very
interesting case. Now we have $a=b=0, \phi=0, E=0$ and so $Z=0$.
The reduction of the equations of motions is simple: just let
$a=0$ and $Z=0$ in the above discussions. And the solutions of the
equations and the metric components could be obtained by setting
$\k_1=\k_2=0$ and $a=0$. We will omit the details here. The answer
looks simpler. Imposing the junction conditions lead to the same
conclusion, namely we have well-defined physical pure-flux-CPW
solutions: a two-parameter family of the $(pqrw)$-type solutions
labelled by $\k_3,\k_4$ and a three-parameter family of the $(f\pm
g )$-type solutions labelled by $\k_3,\k_4$ and $c_1$. Similarly,
the discussion on the future singularity follows directly and
reach the same conclusion.

There are two remarkable solutions in the pure flux case without
dilaton. Firstly, we can have physical CPW solutions in the
higher-dimensional Maxwell-Einstein gravity. Now the flux is a two
form, being the field strength of the $U(1)$ gauge field. So $n=1$
but we have no constraints on the number $m,l$ of other
dimensions. Therefore with metric (\ref{metric}),  there exist not
only the physically well-defined higher-dimensional Bell-Szekeres
$(pqrw)$-type solutions (see also \cite{Gurses3}) but also
well-defined $(f\pm g)$-type solutions. When $m+l=1$, the
$(pqrw)$-type solution reduce to the well-known Bell-Szekeres
solution in four dimension. Secondly, we can have flux-CPW
solutions in the eleven-dimensional supergravity theory which has
no dilaton. This corresponds to $n=2, m+l=7$. In \cite{chen}, with
the metric ansatz (\ref{metric1}), we noticed that the above two
flux-CPW solutions were impossible and the only pure flux-CPW
solutions was four-dimensional BS solution. Now we show that with
the metric ansatz (\ref{metric}), the higher-dimensional flux-CPW
solutions without dilaton are possible. This indicate that in
higher-dimensional gravity theory, the CPW solutions have more
rich structure.

\subsection{Pure Dilatonic CPW solutions}

If we set $a=b=0, D=E=0$, we have the pure dilatonic gravity
theory. After the reduction of the equations of motions, we have
$X=-V, Y=W$ and $Z=\phi$ and three decoupled equations
 \bea
 (f+g)X_{fg}+\frac{1}{2}(X_f+X_g)&=&0 \nonumber \\
 (f+g)Y_{fg}+\frac{1}{2}(Y_f+Y_g)&=&0 \nonumber \\
 (f+g)Z_{fg}+\frac{1}{2}(Z_f+Z_g)&=&0. \nonumber
 \eea
For simplicity, we choose the solutions to the above equations to
be of the Khan-Penrose-Szekeres type, i.e.
 \bea
 X&=&\kappa_3\log\frac{w-p}{w+p}+\kappa_4\log\frac{r-q}{r+q},
 \nonumber \\
 Y&=&\kappa_5\log\frac{w-p}{w+p}+\kappa_6\log\frac{r-q}{r+q},
 \nonumber \\
 Z&=&\kappa_1\log\frac{w-p}{w+p}+\kappa_2\log\frac{r-q}{r+q}.
 \nonumber
 \eea
Then the metric components and the dilaton field are of the forms:
 \bea
 e^{-M}&=&f_{u}g_{v}
\left[(1-2f)(1+2g)\right]^{-b_1}\left[(1+2f)(1-2g)\right]^{-b_2}(f+g)^{-b_3-1+\delta}\nonumber\\
& &\times\left[1+4fg+4pqrw\right]^{-b_4}
\exp\{-\frac{X}{2n}-\frac{Y}{2m}\}
 \label{M4}\\
e^{nA}&=&(f+g) e^{-X}\label{A4}\\
e^{mB}&=&(f+g)e^{-Y} \label{B4}\\
 e^{lC}&=&e^{X+Y}\label{C4}\\
 e^\phi &=&e^Z
 \eea
 where
 \bea
 b_1&=&\frac{\k_3^2}{4n}+\frac{\k_5^2}{4m}+\frac{(\k_3+\k_5)^2}{4l}+\k_1^2
 \\
 b_2&=&\frac{\k_4^2}{4n}+\frac{\k_6^2}{4m}+\frac{(\k_4+\k_6)^2}{4l}+\k_2^2
 \\
 b_3&=&-\left(\frac{(\k_3+\k_4)^2}{4n}+\frac{(\k_5+\k_6)^2}{4m}+\frac{(\k_3+\k_4+\k_5+\k_6)^2}{4l}+(\k_1+\k_2)^2\right)
 \\
 b_4&=&\frac{\k_3\k_4}{2n}+\frac{\k_5\k_6}{2m}+\frac{(\k_3+\k_5)(\k_4+\k_6)}{2l}+2\k_1\k_2
 \eea
 Now we have a six-parameter family of the solutions labelled by
 $\k_i, i=1, \cdots, 6$.

 As before, the junction conditions require that $1\geq b_i \geq
 1/2, i=1,2$, which could be easily achieved. Also the metric is
 singular at $S_0$. And the curvature invariants have singular
 terms $(f+g)^{2\beta}$ near $S_0$, where
  \bea
  \beta&=&b_3-1-\delta+\frac{\k_3+\k_4}{2n}+\frac{\k_5+\k_6}{2m}
  \nonumber \\
  &=&-1-\frac{1}{4n}(\k_3+\k_4-1)^2-\frac{1}{4m}(\k_5+\k_6-1)^2 <0
  \eea
Therefore, it is impossible for the solution to avoid the
singularity in the future.

In this case, if we set $\phi=0$, we have the pure gravitational
CPW solution, which could be reduced from the above solutions in a
direct way. In \cite{Gurses2}, the higher even dimensional pure
CPW solutions have been constructed. The solutions take a quite
similar form as (\ref{M4}). Our solutions exist in any higher
dimensions.

\subsection{Two-flux CPW solutions without dilaton}

If we  set $a=b=0, \phi=0$ such that the dilaton field is turned
off, we obtain a gravity theory without the dilaton. Now we have
$X=-V, Y=W$ and $Z=0$. The equations on $X,Y,D,E$ take the same
form. Therefore, the solutions could be read from
(\ref{M1}-\ref{C1}) and (\ref{M2}-\ref{C2}) directly by setting
$a=b=0, \k_1=\k_2=0$. The imposition of the junction conditions
and the discussion of the future singularity follow in a
straightforward way.



\section{Discussions and Conclusion}

In this paper, starting from the metric (\ref{metric}), we tried
to construct the colliding plane wave solutions in a higher
dimensional gravity theory with the dilaton and the fluxes. We
find that in general we have two classes of well-defined physical
solutions after imposing the junction conditions. Of particular
interest is the one-flux case without the dilaton field. In this
case, we obtain the higher dimensional Bell-Szekeres solutions in
Maxwell-Einstein gravity and also flux-CPW solutions in 11D
supergravity theory.  Our discovery of the well-defined pure
flux-CPW solutions in higher dimension indicate that the flux-CPW
solutions is more intricate since there are more degrees of
freedom in the metric. One may try to find more CPW solutions from
a more generic metric ansatz.

One interesting point is that in our discussion we limit ourselves
to two restricted classes of solutions. It must be possible to
consider other kinds of solutions in our context.

Another point is that the initial background is important to the
formation of the singularity. So naively one can try to study the
CPW solutions in AdS or dS spacetime. In the Anti-de Sitter
spacetime, the study may shed light on the AdS/CFT correspondence.
And in the de Sitter spacetime, the study may have some
cosmological implications, if there do exist a positive tiny
cosmological constant.

\section*{Acknowledgements}

This work was supported by a grant of Chinese Academy of Science.

\appendix
\section{Riemann and  Ricci  tensors}

With the metric ansatz (\ref{metric}), We have Ricci tensor

\begin{eqnarray}
R_{uu}&=&-\frac{1}{2}\left[nA_{uu}+mB_{uu}+lC_{uu}+nM_uA_u+mM_uB_u+lM_uC_u+\frac{1}{2}(nA_u^2+mB_u^2+lC_u^2)\right]\\
R_{vv}&=&-\frac{1}{2}\left[nA_{vv}+mB_{vv}+lC_{vv}+nM_vA_v+mM_vB_v+lM_vC_v+\frac{1}{2}(nA_v^2+mB_v^2+lC_v^2)\right]\\
R_{uv}&=&M_{uv}-\frac{n}{2}A_{uv}-\frac{m}{2}B_{uv}-\frac{l}{2}C_{uv}-\frac{1}{4}(nA_uA_v+mB_uB_v+lC_uC_v)\\
R_{xx}&=&-\frac{1}{2}e^{M+A}\left[2A_{uv}+nA_uA_v+\frac{m}{2}(A_uB_v+A_vB_u)+\frac{l}{2}(A_uC_v+A_vC_u)\right]\\
R_{yy}&=&-\frac{1}{2}e^{M+B}\left[2B_{uv}+mB_uB_v+\frac{n}{2}(A_uB_v+A_vB_u)+\frac{l}{2}(B_uC_v+B_vC_u)\right]\\
R_{zz}&=&-\frac{1}{2}e^{M+C}\left[2C_{uv}+lB_uB_v+\frac{n}{2}(A_uC_v+A_vC_u)+\frac{m}{2}(B_uC_v+B_vC_u)\right]
\end{eqnarray}
where $x = x_i$ with $i = 1, \cdots, n$,  $y = y_j$ with $j = 1,
\cdots,m$ and $z = z_k$ with $k = 1, \cdots, l$. And also we have
the independent non-vanishing components of the Riemann tensor as
following:

\begin{eqnarray}
R_{uvuv}&=&-e^MM_{uv}\\
R_{uxvx}&=&-e^A(\frac{1}{2}A_{uv}+\frac{1}{4}A_uA_v)\\
R_{vxvx}&=&-e^A(\frac{1}{2}A_{vv}+\frac{1}{2}M_vA_v+\frac{1}{4}A_v^2)\\
R_{uxux}&=&-e^A(\frac{1}{2}A_{uu}+\frac{1}{2}M_uA_u+\frac{1}{4}A_u^2)\\
R_{uyvy}&=&-e^B(\frac{1}{2}B_{uv}+\frac{1}{4}B_uB_v)\\
R_{vyvy}&=&-e^B(\frac{1}{2}B_{vv}+\frac{1}{2}M_vB_v+\frac{1}{4}B_v^2)\\
R_{uyuy}&=&-e^B(\frac{1}{2}B_{uu}+\frac{1}{2}M_uB_u+\frac{1}{4}B_u^2)\\
R_{uzvz}&=&-e^C(\frac{1}{2}C_{uv}+\frac{1}{4}C_uC_v)\\
R_{uzuz}&=&-e^C(\frac{1}{2}C_{uu}+\frac{1}{2}M_uC_u+\frac{1}{4}C_u^2)\\
R_{vzvz}&=&-e^C(\frac{1}{2}C_{vv}+\frac{1}{2}M_vC_v+\frac{1}{4}C_v^2)\\
R_{xyxy}&=&-\frac{1}{4}e^{M+A+B}(A_uB_v+A_vB_u)\\
R_{xzxz}&=& -\frac{1}{4}e^{M+A+C}(A_uC_v+A_vC_u)\\
R_{yzyz}&=&-\frac{1}{4}e^{M+B+C}(C_uB_v+C_vB_u)\\
R_{x_ix_{i'}x_ix_{i'}}&=&-\frac{1}{2}e^{M+2A}A_uA_v, \hspace{5ex}
(i\neq i') \\
R_{y_jy_{j'}y_jy_{j'}}&=&-\frac{1}{2}e^{M+2B}B_uB_v, \hspace{5ex}
(j\neq j')\\
R_{z_kz_{k'}z_kz_{k'}}&=&-\frac{1}{2}e^{M+2C}C_uC_v, \hspace{5ex}
(k\neq k').
\end{eqnarray}

Also, we have
 \bea
 R_4&=&
 \frac{e^{2M}}{4}\left(16M^2_{uv}+4n(A_{uu}A_v^2+A_{vv}A_u^2+M_uA_uA_v^2+M_vA_vA_u^2)
 \right. \nonumber \\
 & &+4m(B_{uu}B_v^2+B_{vv}B_u^2+M_uB_uB_v^2+M_vB_vB_u^2)+4l(C_{uu}C_v^2+C_{vv}C_u^2+M_uC_uC_v^2+M_vC_vC_u^2)
  \nonumber \\
 &
 &+2nlA_uC_vA_vC_u+2mnA_uB_vA_vB_u+2mlB_uC_vB_vC_u+2m(m+1)B_u^2B_v^2+2n(n+1)A_u^2A_v^2+2l(l+1)C_u^2C_v^2\nonumber
 \\
& &
 +8n(A_{uv}A_uA_v+A_{uv}^2+A_{uu}A_{vv}+M_uA_uA_{vv}+M_vA_vA_{uu}+M_uA_uM_vA_v)
 \nonumber \\
 & &+8m(B_{uv}B_uB_v+B_{uv}^2+B_{uu}B_{vv}+M_uB_uB_{vv}+B_vB_vB_{uu}+M_uB_uM_vB_v)
 \nonumber \\
  & &
 +8l(C_{uv}C_uC_v+C_{uv}^2+C_{uu}C_{vv}+M_uC_uC_{vv}+C_vC_vC_{uu}+M_uC_uM_vC_v)
 \nonumber \\
 & &\left.
 +nm(A_v^2B_u^2+B_v^2A_u^2)+nl(A_v^2C_u^2+C_v^2A_u^2)+ml(C_v^2B_u^2+B_v^2C_u^2)\right)
 \eea

\end{document}